\documentclass[]{vgtc}                          

\ifpdf
  \pdfoutput=1\relax                   
  \usepackage{graphicx}                
  \DeclareGraphicsExtensions{.pdf,.png,.jpg,.jpeg} 
\else
  \usepackage{graphicx}                
  \DeclareGraphicsExtensions{.eps}     
\fi%

\graphicspath{{figures/}{pictures/}{images/}{./}} 

\usepackage{microtype}                 
\PassOptionsToPackage{warn}{textcomp}  
\usepackage{textcomp}                  
\usepackage{mathptmx}                  
\usepackage{times}                     
\usepackage{cite}                      
\usepackage{tabu}                      
\usepackage{booktabs}                  

\usepackage{array}
\usepackage{booktabs}
\usepackage{makecell}
\newcolumntype{C}[1]{>{\centering\arraybackslash}p{#1}}
\newcolumntype{L}[1]{>{\raggedright\arraybackslash}p{#1}}

\usepackage{amssymb}

\usepackage[longend,ruled,vlined,linesnumbered]{algorithm2e}
\usepackage{algorithmic}

\usepackage{enumitem}

\usepackage{graphicx}
\usepackage{color}
\usepackage{xcolor}
\usepackage{subfigure}
\usepackage{url,times,amsmath,cases}
\usepackage{diagbox}
\usepackage{multirow}
\usepackage{booktabs}

\newcommand {\mymarginpar}[1]{\marginpar{#1}}
\renewcommand {\marginpar}[1]{}

\newcommand {\bsec}[2]{\section{#1}
                       \label{sec:#2} }

\newcommand {\bsubsec}[2]{\mymarginpar{sec:#2}
                       \subsection{#1}
                       \label{sec:#2} }

\newcommand {\rsubsec}[1]{\autoref{sec:#1}}

\newcommand {\beq}[1]{
                      \begin{equation}
                      \label{eq:#1} }
\newcommand {\eeq}{\end{equation}}
\newcommand {\beqno}[1]{\begin{eqnarray}
                      \nonumber}

\newcommand {\eeqno}{ && \end{eqnarray}}

\newcommand {\bear}[1]{
                       \begin{eqnarray}
                       \label{eq:#1} }

\newcommand {\bearno}[1]{
                       \begin{eqnarray}
                       \nonumber}

\newcommand {\eear}{\end{eqnarray}}
\newcommand {\eearno}{\end{eqnarray}}

\newcommand {\btab}[1]{
                       \begin{table}
                       \centering
                       \begin{tabular}{#1}}
\newcommand {\etab}[3] {
                       \end{tabular}
                       \caption[#3]{#2}
                       \label{tab:#1}
                       \end{table}
                       \vspace{.1in}}

\newcommand {\btabular}[1]{\begin{center}
                       \begin{tabular}{#1}}
\newcommand {\etabular}{\end{tabular}
                       \end{center}}

\newcommand {\bdefin}[1]{\begin{definition}\label{def:#1}}
\newcommand {\edefin}       {\end{definition}}

\newcommand {\bpro}[1]{\begin{property}
                      \label{pro:#1} }
\newcommand {\epro}   {\end{property}}

\newcommand {\bprop}[1]{\begin{proposition}
                      \label{prop:#1} }
\newcommand {\eprop}       {\end{proposition}}

\newcommand {\blem}[1]{\begin{lemma}
                      \label{lem:#1}}
\newcommand {\elem}   {\end{lemma}}

\newcommand {\bthe}[1]{\begin{theorem}
                      \label{the:#1} }
\newcommand {\ethe}   {\end{theorem}}

\newcommand {\bcor}[1]{\begin{corollary}
                      \label{cor:#1} }
\newcommand {\ecor}   {\end{corollary}}

\newcommand{\hide}[1]{}

\usepackage{array}
\newcolumntype{L}[1]{>{\raggedright\arraybackslash}p{#1}}

\newlength{\vaStemH}
\newlength{\vaImgH}
\newlength{\vaOptRowH}
\newlength{\vaOptH}

\usepackage{caption}
\usepackage{gensymb}

\definecolor{dark_red}{HTML}{8B0000}

\onlineid{1734}

\vgtccategory{Theoretical \& Empirical}

\title{Model Literacy: An Extra Summative Evaluation Factor for Visual Analytics}

\author{%
Lei Xia$^{1}$,
Siyu Wu$^{1}$,
Haodian Li$^{1}$,
Ye Sun$^{1}$,\\
Liang Zhou$^{2}$\thanks{e-mail: zhoulng@pku.edu.cn},
Lei Shi$^{1}$\thanks{e-mail: leishi@buaa.edu.cn}
}

\affiliation{\scriptsize
$^{1}$Beihang University, Beijing, China\\
$^{2}$National Institute of Health Data Science, Peking University, Beijing, China
}

\abstract{Understanding and enhancing visual analytics (VA) performance is important for maximizing their impact.
Existing studies have successfully applied well-established summative evaluation methods from information visualization to the VA context, yet the recent emphasis on an extra data analysis/modeling stage in the VA pipeline poses an additional challenge.
Inspired by the modern concept of visualization literacy, this paper examines \textit{model literacy}, namely users' knowledge of the analysis model used in a VA technique, as an additional factor for VA performance.
Results from a controlled study on the visual analysis of multidimensional data with two dimensionality-reduction models indicate a positive correlation between model-task accuracy and VA-task accuracy. 
The study involves two common dimensionality-reduction models, PCA and t-SNE.
The correlation is stronger for PCA than for t-SNE in the current task design, a pattern consistent with the possibility that VA effectiveness is more closely associated with model literacy when model outputs are less directly readable from the visualization.
Completion-time evidence does not show a stable efficiency gain, suggesting that differences in model intuitiveness may help explain when model knowledge shortens task completion and when it involves additional interpretive effort.
The findings of this study suggest ways to further enrich VA evaluation methods and provide directions for developing more rigorous model-literacy assessment instruments.
} 


\renewcommand{\manuscriptnotetxt}{}

\nocopyrightspace

\vgtcinsertpkg

\begin{document}




\maketitle


\bsec{Introduction}{Intro}


Visual analytics (VA) was originally defined as the science of analytical reasoning facilitated by interactive visual interfaces~\cite{thomas2006visual}.
Today, it is increasingly associated with the integration of visualization and automated analysis techniques, such as data mining and machine learning \cite{keim2008visual}, to enhance human cognitive capabilities in the era of big data.
Over the past decades, the field has grown through new VA techniques for diverse data and models (network, text, spatiotemporal, etc.), expanding application domains (security, biomedicine, social science, etc.), and emerging VA theories~\cite{chen2019ontological,ceneda2017guidance}.


This work focuses on a critical issue of VA techniques: their evaluation, which is a complex topic that sits at the intersection of theoretical and empirical aspects of VA research.
In the literature, a majority of evaluations applied for VA techniques follow classical methods for evaluating InfoVis/SciVis designs, utilizing quantitative experiments, case studies, and researcher inspections \cite{khayat2019validity,scholtz2022user}.
Some specialized VA evaluations measure the number of insights obtained during the VA process \cite{saraiya2005insight,saraiya2006insight}, or they assess the system's sensemaking potential~\cite{kang2012examining}.
Nevertheless, most existing VA evaluation designs are limited to the initial VA definition of ``interactive visualization + analytical reasoning''.
As VA increasingly combines visualization with automated data analysis models, the corresponding model-related evaluation factors have received limited attention.

We aim to identify a key distinction in the summative evaluation of VA techniques under this model-driven paradigm.
Note that we restrict our scope to the summative evaluation of VA systems after they are built, whose functionality is either analyzing advanced data patterns or understanding model output.
The main idea of this work stems from comparing the well-known InfoVis pipeline \cite{card1999readings} with the current practice of VA processes, in which there is an additional stage of data modeling or automated analysis.
While research on introducing statistical or predictive data analysis models to VA design is extensive, there is a lack of studies examining how users' knowledge about these models is related to VA performance.
The potential mismatch between users' modeling expertise and complex model-driven VA systems motivates our work.

Inspired by the established concept of visualization literacy, we introduce \textit{model literacy} as an additional factor in VA evaluation.
We refer to it as a user's knowledge of a data analysis model in the VA pipeline, including how the model describes or predicts data and how its input and output should be interpreted.
In this paper, a data analysis model refers to a mathematical or computational component that transforms input data into analytical outputs subsequently represented and interpreted in a VA system.

Unlike a familiar chart whose axes directly encode original variables, a visual representation of model output in VA may require knowledge of the model mechanism before its visual patterns can be correctly interpreted.
For instance, a user may be able to read a scatterplot but still be unable to reason from the plot if its coordinates are defined by principal components or a nonlinear embedding unknown to the user.
Because the embedded model determines the analytical quantities represented in the visualization, misunderstanding these quantities may cause an evaluation to attribute model-related interpretation errors to visual encoding, interaction, or interface design.

The main contributions of this work are as follows:
\begin{itemize}
  \item We introduce model literacy as an extra factor for the summative evaluation of VA systems. 
  This factor captures users' knowledge of the data analysis model embedded in the VA pipeline and complements existing evaluation concerns centered on visualization, interaction, and analytical reasoning.
  \item We formulate tests of model literacy in a controlled VA evaluation setting by separating visualization tasks, model tasks, and VA tasks. 
  This design allows us to examine the relation between model-task accuracy and VA performance while reducing variation 
  attributable to data familiarity and basic scatterplot reading.
  \item We provide empirical evidence from a user study on multidimensional data analysis using PCA and t-SNE as two dimensionality-reduction conditions. 
  Model-task accuracy is positively associated with VA-task accuracy under both models, with a stronger association under PCA that is consistent with a possible role of relative model intuitiveness in the current task design.
\end{itemize}

\bsec{Related Work}{Related}
Our work is related to summative visualization evaluation, evaluation practices for visual analytics, and visualization literacy and related literacies in visual analytics.

\bsubsec{Summative Evaluation Methods for Visualization}{RelatedVisEval}

Evaluation has been an important topic for visualization ever since the birth of the field~\cite{plaisant2004challenge,isenberg2013systematic}.
Based on the usage stage within the visualization life cycle, evaluation methods can be categorized into explorative, predictive, formative, and summative~\cite{andrews2008evaluation}, although the classification boundary can be blurred in some cases due to their different evaluation intentions~\cite{khayat2019validity}.


In the field of visualization, summative evaluation is originally defined as a method to assess the usefulness of visualization designs/systems after they are implemented.
A quantitative lab experiment is a typical way to conduct summative evaluation of visualizations.
To this end, many well-developed evaluation theories for visualization can also serve the summative purpose.
For example, Lam et al. identify seven evaluation scenarios in information visualization, which include three for understanding visualization: evaluating user performance, user experience, and visualization algorithms~\cite{lam2011empirical}.
This scenario-based view is useful here because it clarifies that different evaluation claims require different forms of evidence.
The BELIV workshop has been organized to further advance evaluation methods beyond the classical quantitative user experiments~\cite{bertini2008beliv}.
Recent BELIV work also emphasizes that evaluation strategies should be aligned with the contribution type and research goal, and that task misunderstandings can threaten the validity of visualization experiments~\cite{lin2024striking,sarma2024tasks}.
Notably, Shneiderman et al. propose multi-dimensional, in-depth, long-term case studies to advocate longitudinal evaluation practice, which can potentially revamp summative evaluation results~\cite{shneiderman2006strategies}.
In addition, the famous nested model~\cite{munzner2009nested}, although mostly formative/predictive, provides several guidelines and techniques to summatively validate the usefulness of visualization-related components at each level of the nested model.

While numerous evaluation methods are available for visualization, most focus on visualization-centric factors, such as visual designs, representations, and interactions.
The impact of model-related user factors on VA performance has not been sufficiently discussed yet.

\bsubsec{Evaluation Practices for Visual Analytics}{RelatedVAEval}



As VA was originally defined as analytical reasoning facilitated by interactive visual interfaces, most VA evaluations still focus on visualization and reasoning factors~\cite{khayat2019validity,blascheck2015va}.
Khayat et al. provide a taxonomy of methods adopted in VAST papers~\cite{khayat2019validity}; most use quantitative experiments, case studies, or inspections to assess visualization and analytical performance.
Lam et al. likewise identify visual data analysis and reasoning as a distinct evaluation scenario~\cite{lam2011empirical}.
More operationally, Amar and Stasko identify two types of analytic gaps in applying interactive visualization for analytical reasoning and formulate knowledge tasks for bridging them~\cite{amar2005knowledge}.
These tasks support goals such as domain learning and complex decision-making, which are closely related to insight-obtaining and sensemaking in VA.


Insight has been defined as complex, qualitative, unexpected, and relevant understanding obtained from data~\cite{north2006toward}.
Insight-based evaluation uses such understanding as a measure of VA effectiveness and combines qualitative observation with quantitative metrics, notably insight counts~\cite{saraiya2005insight,saraiya2006insight}.
Compared with task accuracy and completion time alone, this approach captures a broader range of analytical outcomes over a longer analysis process.
Sensemaking is harder to observe directly and is therefore often evaluated through case studies~\cite{kang2012examining}.
Pair analytics combines a subject-matter expert and a VA expert in a collaborative setting to capture the sensemaking process~\cite{arias2011pair}.
It also addresses limitations of individual protocol analysis, including limited subject expertise and cognitive stamina.


The interdisciplinary nature of VA has also motivated other evaluation approaches.
Chen and Ebert propose an ontological framework for diagnosing problems and potential causes in VA systems~\cite{chen2019ontological}, supported by a cost--benefit view of VA workflows~\cite{chen2015may}.
For investigative analysis, comparative evaluation contrasts VA systems with traditional methods~\cite{kang2009evaluating}.
Both quantitative usage patterns and verbal feedback can contribute to the comparison.
Recent BELIV work on micro-entries further shows that users' mental models of interactive data systems can evolve during interaction and may require deeper evaluation~\cite{block2020microentries}.
This line of work is close to our concern with users' understanding of system logic, but our focus is narrower: users' knowledge of the data analysis model embedded in the VA pipeline and its relation to summative VA performance.


In summary, most VA evaluation methods still follow the classical ``visualization + reasoning'' view.
Model-driven VA, which combines data analysis models with interactive visualization for knowledge generation, introduces additional evaluation needs~\cite{keim2008visual,sacha2014knowledge}.
Although algorithms have been identified as an evaluation component~\cite{scholtz2022user}, this discussion mainly concerns system or visualization algorithms rather than users' understanding of data analysis models.
Prior work has also examined roadblocks for visualization novices~\cite{chul2011visual}, but evaluation for users with limited understanding of embedded models remains scarce.

\bsubsec{Visualization Literacy and Related Literacies}{RelatedLite}

Visualization literacy is a measurable user factor associated with visualization effectiveness.
Early work examined graph comprehension and graphical perception, focusing on how people read charts, decode visual encodings, and derive meaning from graphical representations~\cite{cleveland1984graphical,carpenter1998graphcomprehension}.
Later studies developed graph-literacy measures~\cite{galesic2011graphliteracy} and assessment instruments such as VLAT~\cite{lee2017vlat}.
The construct has since expanded beyond reading and interpretation toward broader competence in making sense of patterns and constructing visualizations~\cite{borner2019dvlfw,ge2025avec}.

Existing visualization-literacy research spans three main themes.
Assessment work ranges from foundational and shortened instruments such as VLAT and Mini-VLAT~\cite{lee2017vlat,pandey2023minivlat} to construction-oriented measures~\cite{ge2025avec} and tests with domain experts~\cite{oney2024testing}.
Interpretation research examines how novices understand unfamiliar visualizations and what barriers interrupt this process~\cite{lee2016novis,nobre2024barriers}.
The NOVIS model frames novice understanding as a staged sensemaking process~\cite{lee2016novis}, while recent work also relates literacy differences to visual inspection behavior~\cite{chang2026tell}.
Education and intervention studies explore elementary, game-based, and construction-oriented support~\cite{chevalier2018elementary,adelberger2025iguanodon}.
Across these themes, literacy primarily concerns perceiving, reading, interpreting, or constructing visual representations; understanding the analytical models behind them has received less explicit attention.

Recent work notes that visualization literacy is not always consistently operationalized and that measurement can vary with task design and context~\cite{lily2025vislit}.
PREVis examines perceived readability as a related but distinct construct~\cite{previs2025}, while multiliteracy perspectives situate interactive visualization within multiple interacting literacies~\cite{multiliteracy2026}.Related concepts such as algorithm literacy and AI literacy likewise concern different objects, ranging from algorithmic systems to broader AI technologies~\cite{AlgorithmLite,AILite}.

A gap nevertheless remains in model-driven VA, where users interpret visual representations of model-generated analytical results rather than raw data alone~\cite{keim2008visual,sacha2014knowledge}.
Correct interpretation may require knowledge of the underlying model---for example, how PCA projects samples or how t-SNE embeddings should be interpreted---that conventional visualization-literacy assessments may not explicitly capture.
This gap motivates our study of model literacy in visual analytics.


\begin{figure*}[t]
\centering
\includegraphics[width=\textwidth, clip, trim=20pt 20pt 20pt 20pt]{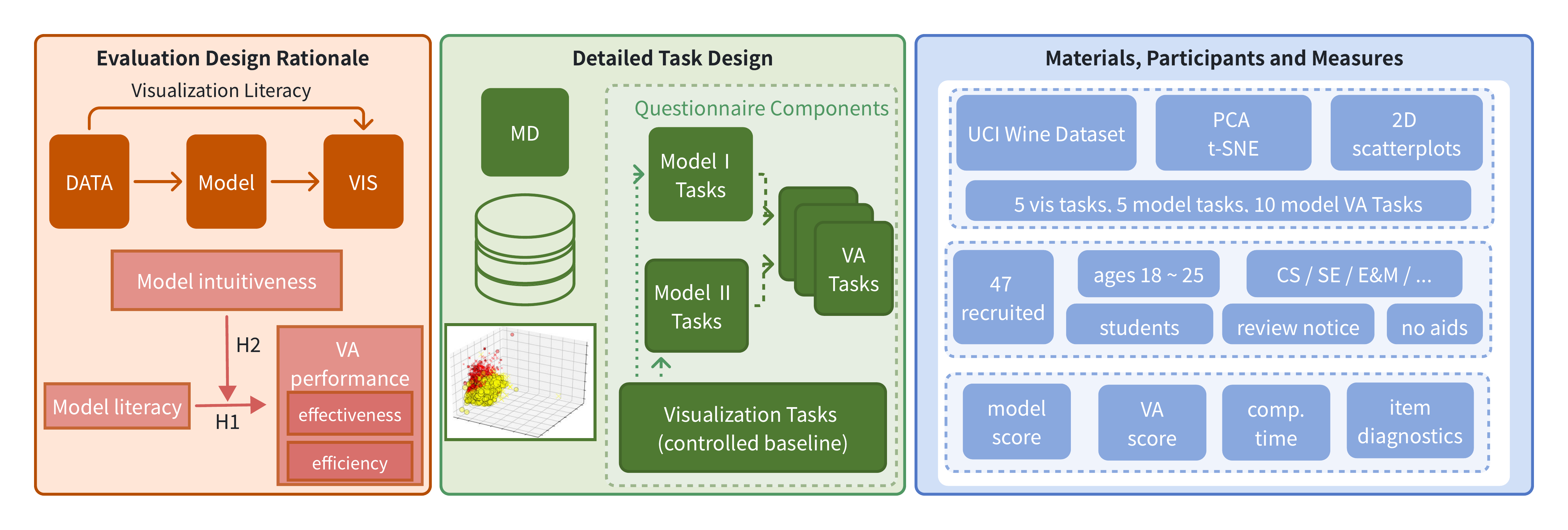}
\caption{Overview of the evaluation design.
The left panel shows the conceptual relations among model literacy, model intuitiveness, and VA performance through hypotheses H1--H2.
The middle panel summarizes the questionnaire components: visualization tasks as a controlled baseline, model tasks for measuring model literacy through model-task accuracy, and VA tasks for measuring VA performance through VA-task accuracy and mean VA-task completion time.
The right panel previews the concrete multidimensional-data instantiation, including the UCI Wine dataset, PCA and t-SNE models, two-dimensional scatterplots, recruited participants, the pre-questionnaire preparation notice, and the measurements used in the study.}
\label{fig:pipeline}
\end{figure*}


\bsec{The Model Literacy Factor and Evaluation Design}{Method}

An overview of our study is shown in Figure~\ref{fig:pipeline}, which summarizes the conceptual motivation, questionnaire framework, study materials, participant pool, and measurements of this work.

\bsubsec{Definition and Hypotheses}{Def}

Our work is motivated by the observation that state-of-the-art VA techniques have significantly evolved from the classical InfoVis pipeline.
According to the conceptual framework by Keim et al., a model is recognized as a first-class component of the data-to-knowledge VA process~\cite{keim2008visual}.
In comparison, the InfoVis pipeline only includes data transformation as a transitional operator~\cite{card1999readings}.
Here, models include descriptive and predictive mathematical procedures and therefore cover a broader scope than elementary data transformation.
Moreover, the additional model component has important implications for the user-centered evaluation of VA techniques.
For example, a model-generated scatterplot may require users to understand how the plotted coordinates are produced before its visual patterns can be correctly interpreted.

Both the previous example and the conceptual rationale motivate examining an additional evaluation factor for VA techniques.
Previous work has shown that the usefulness of visualization is influenced by both design excellence and the user's visualization literacy~\cite{lee2017vlat}.
Analogously, model literacy can be an extra factor for VA techniques using a data analysis model, particularly when users make sense of data through visual representations of model output.

We define model literacy as \emph{a user's knowledge and competence in understanding how a data analysis model works, how its output should be interpreted, and how model-generated results can be used in visual analytics}.
Specifically, the literacy involves three types of understanding:
1) understanding the mapping mechanism between the input and output data of the model, including the advanced data transformation mechanism for descriptive models and the prediction rationale for predictive models;
2) correctly comprehending and analyzing the output of a model for data patterns;
3) appropriately evaluating, selecting, configuring, and applying a data analysis model.

In this work, because we focus on the summative evaluation of VA systems after their implementation, we focus on the first two parts of the definition related to model comprehension and output interpretation.
Evaluating the last component, which concerns model design and usage, can be a promising direction for future work.
In the literature, there are also related definitions such as algorithm literacy~\cite{AlgorithmLite} and AI literacy~\cite{AILite}.
While AI literacy ostensibly covers a much broader scope across all AI technologies, algorithm literacy comes closer to our study.
Nevertheless, algorithm literacy is defined to promote the user's awareness and knowledge of algorithmic decision-making on the Internet, which may not always be linked to the analysis of data.
In contrast, our definition focuses on the user's knowledge of data analysis models and model-generated results in visual data analysis rather than broader awareness of algorithmic decision-making.

We formulate two hypotheses on the relations among model literacy, model intuitiveness, and VA performance, as shown in Figure~\ref{fig:pipeline}-left.
We use \emph{VA performance} as an umbrella term for VA effectiveness and VA efficiency.
VA effectiveness concerns whether users make correct downstream VA judgments, whereas VA efficiency concerns how much time is required to reach those judgments.
In this work, we use \textit{more intuitive models} to refer to data analysis models whose input--output mappings are relatively direct and expected for users with basic relevant knowledge.
In the multidimensional-data experiment, PCA represents a relatively less intuitive model and t-SNE a relatively more intuitive model: t-SNE preserves local neighborhoods in the embedding and therefore better matches common spatial judgments about high-dimensional similarity, whereas PCA requires users to interpret linear components and feature loadings.

\textbf{H1:} Users with more knowledge of a specific data analysis model, i.e., higher model literacy, achieve higher VA effectiveness (higher task accuracy) and higher VA efficiency (shorter task completion time).

\textbf{H2:} The correlation between model literacy and VA performance is stronger for relatively less intuitive models and weaker for more intuitive models.

\bsubsec{Overall Evaluation Design}{Task_Design}

We design the experiment to evaluate the VA process composed of three key stages, as shown in Figure~\ref{fig:pipeline}-left.
The main rationale is to reduce variation induced by data familiarity and basic scatterplot reading.
We do so by selecting an entry-level data type and a visual representation familiar to the participant pool.
This allows VA performance to be examined in relation to model-task accuracy as an operational measurement of model literacy.

The questionnaire framework is organized around three task types (Figure~\ref{fig:pipeline}-middle).
Visualization tasks serve as a controlled baseline for whether participants can read the visual representation itself.
Model tasks measure model literacy through model-task accuracy.
VA tasks then examine whether participants can use the designed views to complete the intended analysis.
We measure VA effectiveness by VA-task accuracy and VA efficiency by mean VA-task completion time, while treating completion time cautiously as both an efficiency indicator and a process-related measure.

We instantiate the evaluation on multidimensional data and use a simple and familiar visual representation: two-dimensional scatterplots.
This design reduces unnecessary variation of the visual representation itself and keeps the study focused on the relationship among model literacy, model intuitiveness, and VA performance.
The concrete questionnaire blocks, task order, data/model materials, generation settings, and representative items are reported at the beginning of Sect.~\ref{sec:Exp}, followed by the corresponding results.

\begin{figure}[t]
\centering
\scriptsize
\textbf{PCA}\par
\vspace{0.5mm}
\parbox{\linewidth}{Given the loadings, which relation among $X_1,X_2,X_3$ and $Y_1,Y_2,Y_3$ cannot hold if all other features are identical?}\par
\vspace{0.5mm}
\begin{minipage}[c]{0.46\linewidth}
\centering
\includegraphics[width=\linewidth]{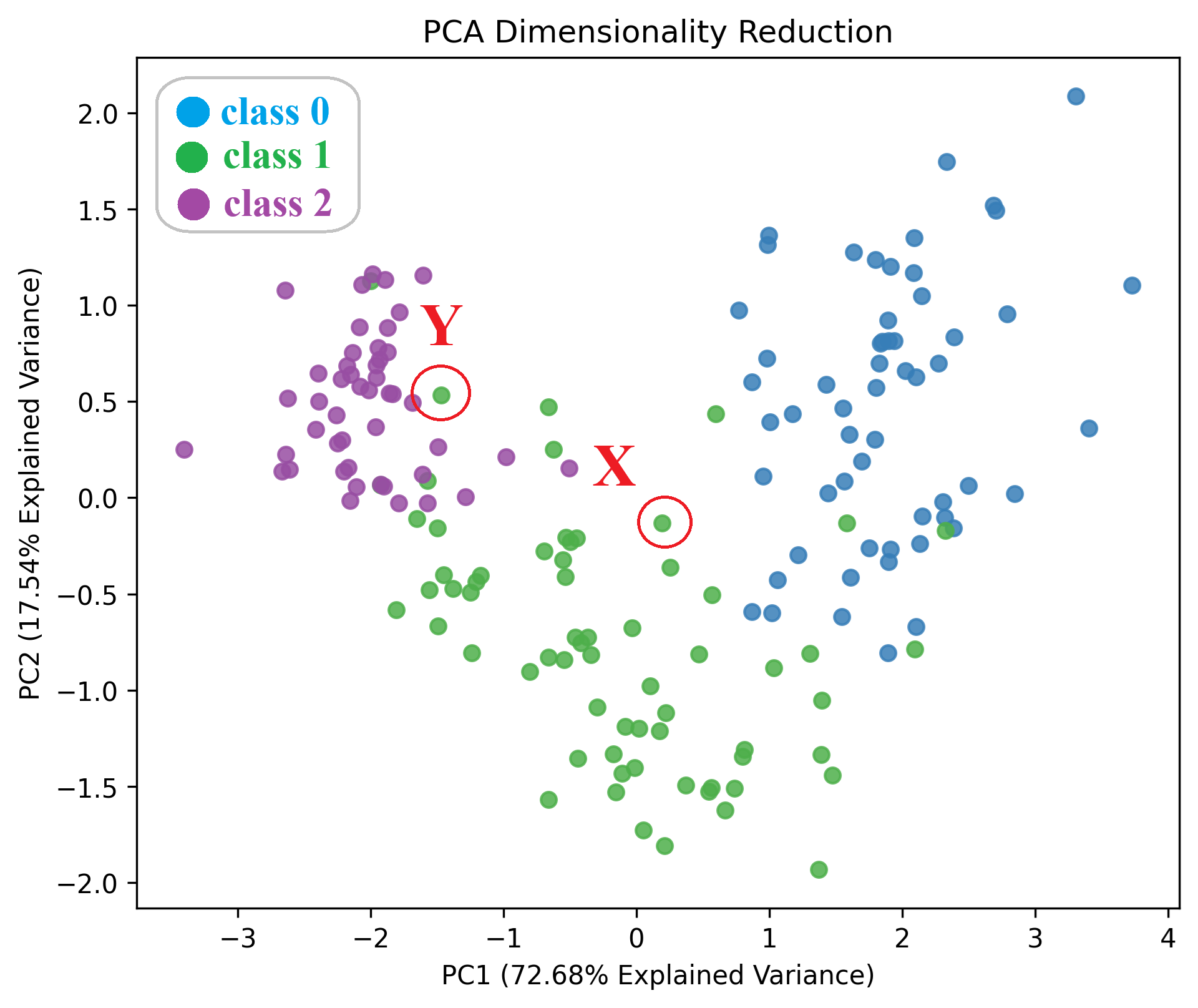}
\end{minipage}\hfill
\begin{minipage}[c]{0.50\linewidth}
\begin{tabular}{@{}p{0.10\linewidth}p{0.84\linewidth}@{}}
A. & $X_1<Y_1,\;X_2>Y_2,\;X_3=Y_3$ \\
B. & $X_1>Y_1,\;X_2<Y_2,\;X_3=Y_3$ \\
C. & $X_1<Y_1,\;X_2>Y_2,\;X_3<Y_3$ \\
D. & $X_1>Y_1,\;X_2<Y_2,\;X_3>Y_3$ \\
\end{tabular}
\end{minipage}

\vspace{1.5mm}
\textbf{t-SNE}\par
\vspace{0.5mm}
\parbox{\linewidth}{Based on the relative positions of the two marked wine samples, which conclusion is the most justified?}\par
\vspace{0.5mm}
\begin{minipage}[c]{0.46\linewidth}
\centering
\includegraphics[width=\linewidth]{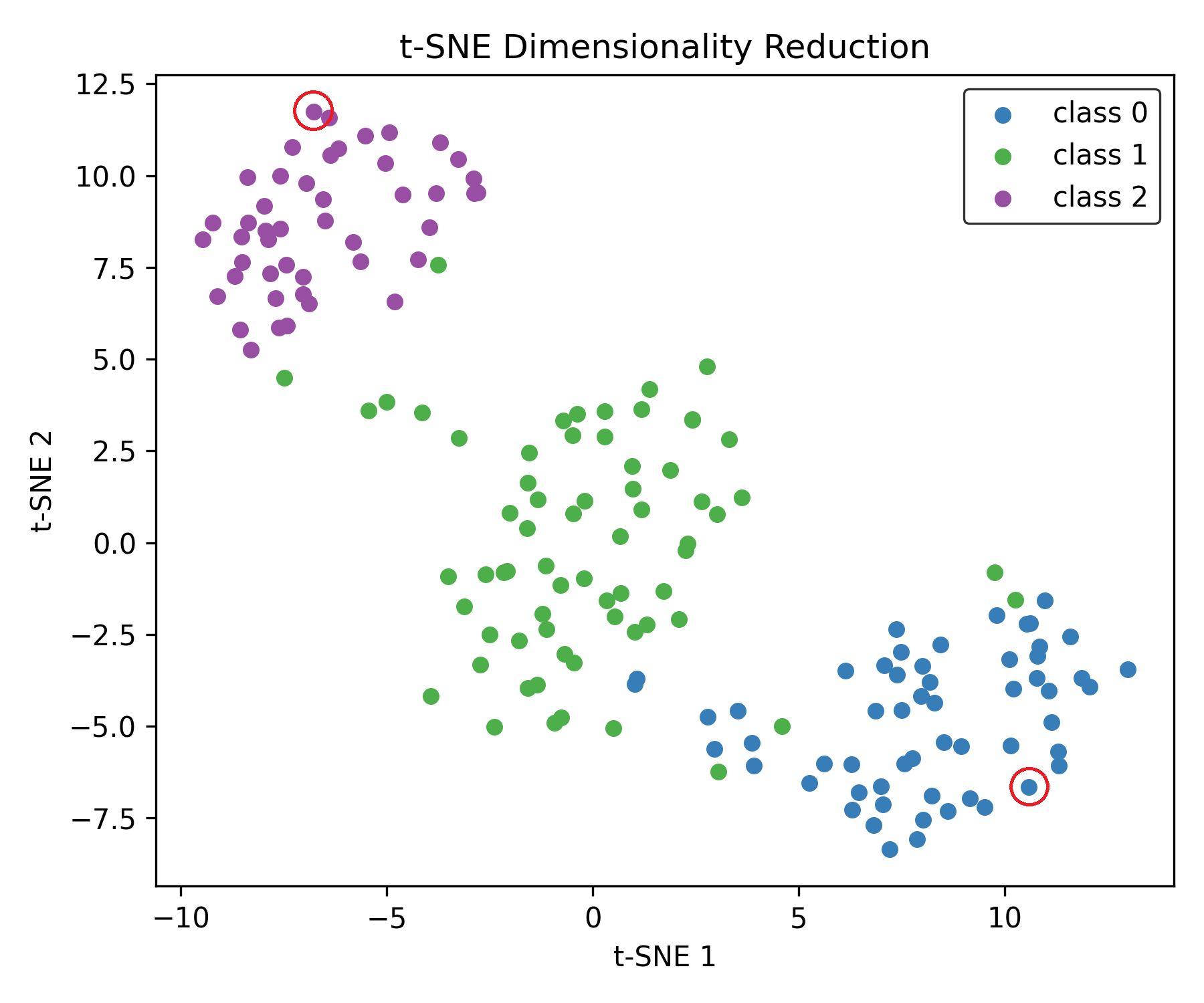}
\end{minipage}\hfill
\begin{minipage}[c]{0.50\linewidth}
\begin{tabular}{@{}p{0.10\linewidth}p{0.84\linewidth}@{}}
A. & Their distance is clearly above average. \\
B. & Their distance is about average. \\
C. & Their distance is clearly below average. \\
D. & Their distance cannot be determined. \\
\end{tabular}
\end{minipage}
\caption{Example VA tasks used in the multidimensional-data questionnaire.}
\label{fig:va_examples}
\end{figure}

\bsubsec{Materials, Participants, and Procedures}{Procedure}

The right panel of Figure~\ref{fig:pipeline} summarizes the concrete multidimensional-data instantiation of the framework.
The study uses the UCI Wine dataset, PCA and t-SNE as the two data analysis models, and two-dimensional scatterplots as the shared visual representation.
For each model, the questionnaire includes model tasks and VA tasks. The task composition and principal model settings are summarized in Sect. 4, while the full questionnaire materials are provided in the supplementary material.

We recruited 47 participants through social networks.
The participants were 18--25 years old and were university students from computer science, software engineering, business/economics, biomedical disciplines, and related fields.
A subset of participants received a pre-questionnaire preparation notice indicating that PCA- and t-SNE-related knowledge would be assessed.
They were allowed to review these topics independently, but the study did not prescribe learning materials, review duration, or the interval between review and questionnaire completion.
We therefore do not treat this notice as a controlled learning intervention; it is recorded only as a preparation-notice status and considered in robustness analysis.
During the questionnaire session, all participants were instructed to answer independently and continuously without external aids or tools.

The study measures model literacy through model-task accuracy and VA performance through VA-task accuracy and mean VA-task completion time.
The visualization-task block is used only as a controlled baseline for basic scatterplot reading, rather than as a standalone measurement of general visualization literacy.
In addition, we report item difficulty and corrected item--total correlation following the item-analysis rationale of VLAT and Mini-VLAT~\cite{lee2017vlat,pandey2023minivlat}.
We also report Kuder--Richardson Formula 20 (KR-20)~\cite{kuder1937theory} as a preliminary internal-consistency estimate for dichotomously scored model-task items.
These diagnostics are used as supporting checks for interpreting the current model-task items.


\bsec{Experimental Results}{Exp}

We conducted a user study via an online questionnaire.
The study was approved by the relevant institutional review board, with the protocol identifier omitted for double-blind review.
All participants provided informed consent before the study began, and each participant was compensated with approximately \$10 USD.
The questionnaire yielded 47 responses. For quality control, we retained responses whose summed item-response time lay between one third and three times the sample median, resulting in 44 retained responses.
This section introduces the concrete questionnaire instantiation, summarizes descriptive performance and item diagnostics, and then reports the empirical evidence for H1 and H2.

The questionnaire contained 35 scored single-choice items on multidimensional data in five blocks: 5 visualization tasks, 5 PCA model tasks, 10 PCA VA tasks, 5 t-SNE model tasks, and 10 t-SNE VA tasks.
The visualization block was presented first as a baseline for scatterplot reading; within each model-specific block, model tasks preceded VA tasks, and two questionnaire versions counterbalanced the order of the PCA and t-SNE blocks.
Model tasks assessed model mechanisms and outputs, whereas VA tasks required applying that knowledge to the displayed scatterplots.
Table~\ref{tab:questionnaire} gives a compact summary of the five task blocks; detailed item coverage, answer options, and answer keys are included in the supplementary material.

\begin{table}[t]
\centering
\caption{Composition of the multidimensional-data questionnaire. Detailed item coverage, visual materials, answer options, and answer keys are included in the supplementary material.}
\label{tab:questionnaire}
\scriptsize
\setlength{\tabcolsep}{3pt}
\renewcommand{\arraystretch}{1.05}
\begin{tabular}{lcl}
\toprule
\textbf{Block} & \textbf{Items} & \textbf{Role} \\
\midrule
Visualization tasks & 5 & Basic scatterplot reading \\
PCA model tasks & 5 & PCA mechanisms and outputs \\
PCA VA tasks & 10 & Applying PCA knowledge \\
t-SNE model tasks & 5 & t-SNE mechanisms and outputs \\
t-SNE VA tasks & 10 & Applying t-SNE knowledge \\
\bottomrule
\end{tabular}
\end{table}

The study used the UCI Wine dataset~\cite{uci_wine}, which contains 178 samples, 13 numerical features, and 3 classes.
A fixed 170-sample subset was randomly selected once and used for all questionnaire materials.
Both model conditions used two-dimensional scatterplots.
PCA views used four standardized features and supplied first-two-component loadings when required; t-SNE views used four features with perplexity 30.
Feature subsets were selected for clear layouts and fixed before deployment, with library defaults used for non-targeted parameters.

The bilingual online questionnaire presented one item per page, allowed neither feedback nor response revision, and supplied the dataset, feature, parameter, and auxiliary information needed before each VA block.
Each item had four options and one predefined correct answer.
The questions were developed by the authors and reviewed by two visualization experts; no separate pilot study was conducted.
Figure~\ref{fig:va_examples} shows two representative VA-task items. The PCA item requires participants to combine projected positions with feature loadings, whereas the t-SNE item tests whether participants avoid treating low-dimensional global distance as a direct high-dimensional distance.

Some participants received an advance PCA/t-SNE topic notice before completing the questionnaire and were allowed, but not required, to review these topics independently.
No learning materials, review duration, or interval between receiving the notice and completing the questionnaire were prescribed or recorded.
During the questionnaire session, all participants were instructed to answer independently and continuously, without external aids or tools.
The notice is recorded and adjusted for in a robustness check, but it is not treated as a separate learning intervention.

\bsubsec{Descriptive Performance and Measurement Checks}{Results_Overview}

Table~\ref{tab:descriptive} summarizes accuracy and response time for the five task blocks.
Visualization-task accuracy is near ceiling (mean $=0.973$, median $=1.000$): four of the five items were answered correctly by every retained participant, and the remaining item had an accuracy of $0.864$.
These items therefore function as control checks for elementary scatterplot reading in the current questionnaire.

\begin{table}[t]
\centering
\caption{Descriptive performance for the~44 retained responses.~Per-item completion time~is the participant-level mean response time within the corresponding block.}
\label{tab:descriptive}
\scriptsize
\setlength{\tabcolsep}{2.5pt}
\renewcommand{\arraystretch}{1.05}
\begin{tabular}{lcrr}
\toprule
\textbf{Task block} & \textbf{Items} & \textbf{Accuracy} & \textbf{Time (s)} \\
 & & \textbf{mean (SD)} & \textbf{mean (SD)} \\
\midrule
Visualization & 5 & $0.973\;(0.069)$ & $16.6\;(16.1)$ \\
PCA model & 5 & $0.677\;(0.249)$ & $33.6\;(30.4)$ \\
PCA VA & 10 & $0.577\;(0.214)$ & $67.1\;(39.9)$ \\
t-SNE model & 5 & $0.714\;(0.284)$ & $34.6\;(30.0)$ \\
t-SNE VA & 10 & $0.568\;(0.214)$ & $35.8\;(26.0)$ \\
\bottomrule
\end{tabular}
\end{table}

Performance becomes more variable once model-related reasoning is required.
Mean model-task accuracy is $0.677$ for PCA and $0.714$ for t-SNE, while mean VA-task accuracy is $0.577$ and $0.568$, respectively.
Within each model, model-task accuracy exceeds VA-task accuracy: the mean gap is $0.100$ for PCA and $0.145$ for t-SNE, with $59.1\%$ and $68.2\%$ of participants, respectively, scoring strictly higher on the model tasks (one-sided Wilcoxon tests, both $p<.001$).
This pattern indicates that knowing general model properties and applying that knowledge to a concrete visual judgment are related but non-identical stages.

Across models, mean model-task accuracy ($p=.538$), mean VA-task accuracy ($p=.518$), and per-item model-task completion time ($33.6$~s vs.~$34.6$~s; $p=.986$) are similar.
The largest aggregate timing difference occurs in the VA blocks: participants spend $67.1$~s per PCA VA item and $35.8$~s per t-SNE VA item, a mean paired difference of $31.3$~s ($p<.001$).
This timing difference is retained as descriptive and process-related context for VA performance.

Because the H1 and H2 analyses use the five-item model-task scores, we also examined their item behavior.
Following the item-analysis rationale of VLAT and Mini-VLAT~\cite{lee2017vlat,pandey2023minivlat}, we report item difficulty and corrected item--total correlation (CITC) for the model-task items.
Because the items are dichotomously scored, we additionally report KR-20 as a preliminary internal-consistency estimate. 

\begin{table}[t]
\centering
\caption{Diagnostic statistics for the five-item model-task blocks. Difficulty is the proportion correct. CITC denotes corrected item--total correlation. KR-20 is a preliminary internal-consistency estimate.}
\label{tab:itemquality}
\scriptsize
\begin{tabular}{lcc}
\toprule
 & \textbf{PCA} & \textbf{t-SNE} \\
\midrule
Mean difficulty & 0.677 & 0.714 \\
Mean CITC & 0.209 & 0.371 \\
KR-20 & 0.411 & 0.613 \\
\bottomrule
\end{tabular}
\end{table}

Item difficulty ranges from $0.545$ to $0.841$ for PCA and from $0.614$ to $0.795$ for t-SNE, avoiding obvious floor and ceiling effects.
For PCA, four items have positive but modest CITCs, while the standardization item has a near-zero relation with the remaining questions (CITC range: $-0.010$---$0.295$); the block's KR-20 is $0.411$.
For t-SNE, all CITCs are positive (range: $0.211$---$0.502$), and KR-20 is $0.613$.
These diagnostics indicate that the model-task blocks retain useful variation for the current study and provide a basis for interpreting model-task accuracy as the observed model-literacy measure, while also pointing to future refinement of the item pool.

Taken together, the descriptive results provide the measurement context for the hypothesis analysis: 
basic scatterplot reading is near the ceiling, the model-task scores retain participant-level variation without strong floor or ceiling effects, and PCA and t-SNE show similar mean accuracy. Completion-time differences are retained as descriptive context for process-related VA performance.

\bsubsec{Evidence for H1: Model Literacy and VA Performance}{H1Results}

\begin{figure}[t]
\centering
\includegraphics[width=0.48\columnwidth]{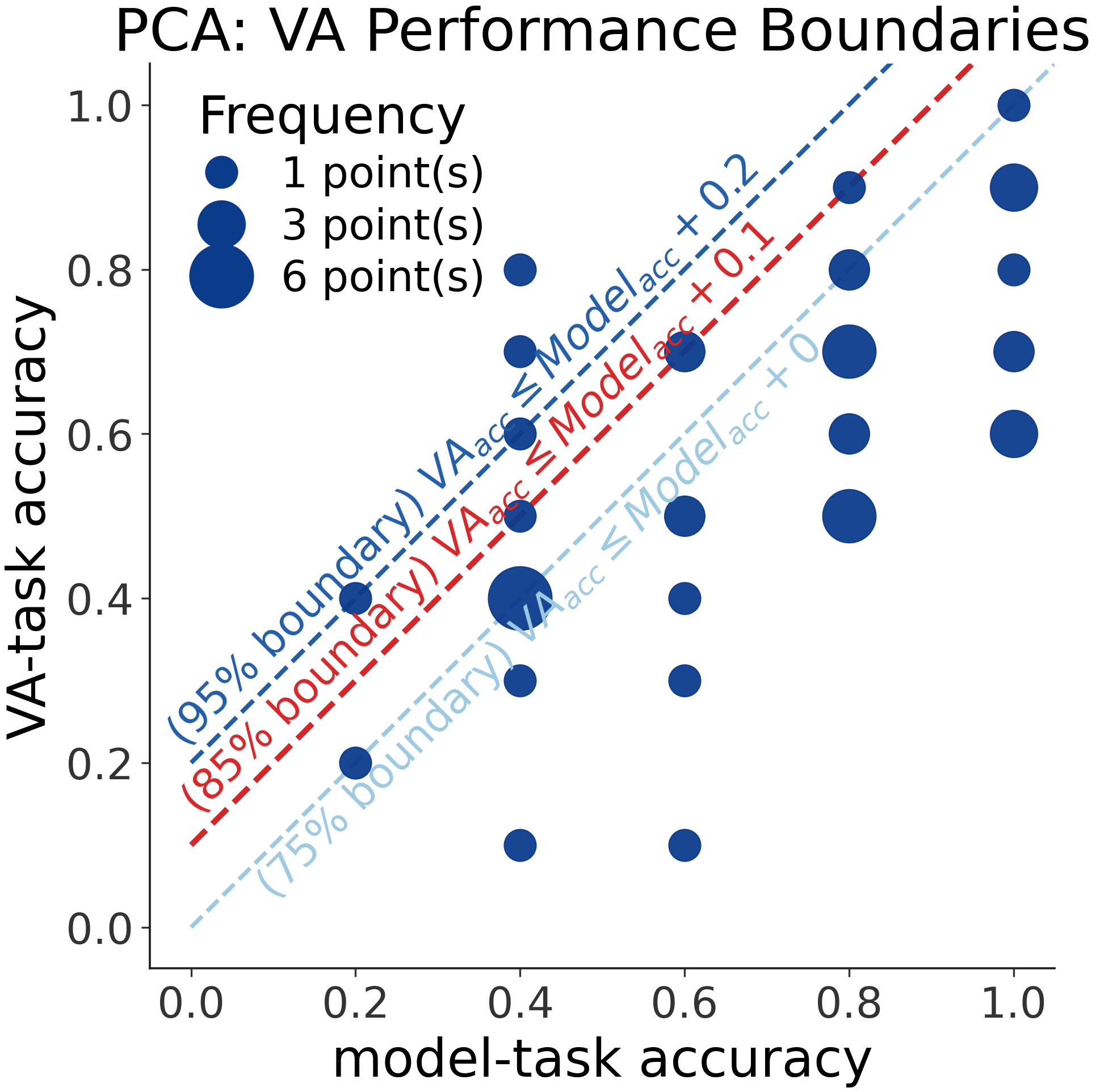}\hfill
\includegraphics[width=0.48\columnwidth]{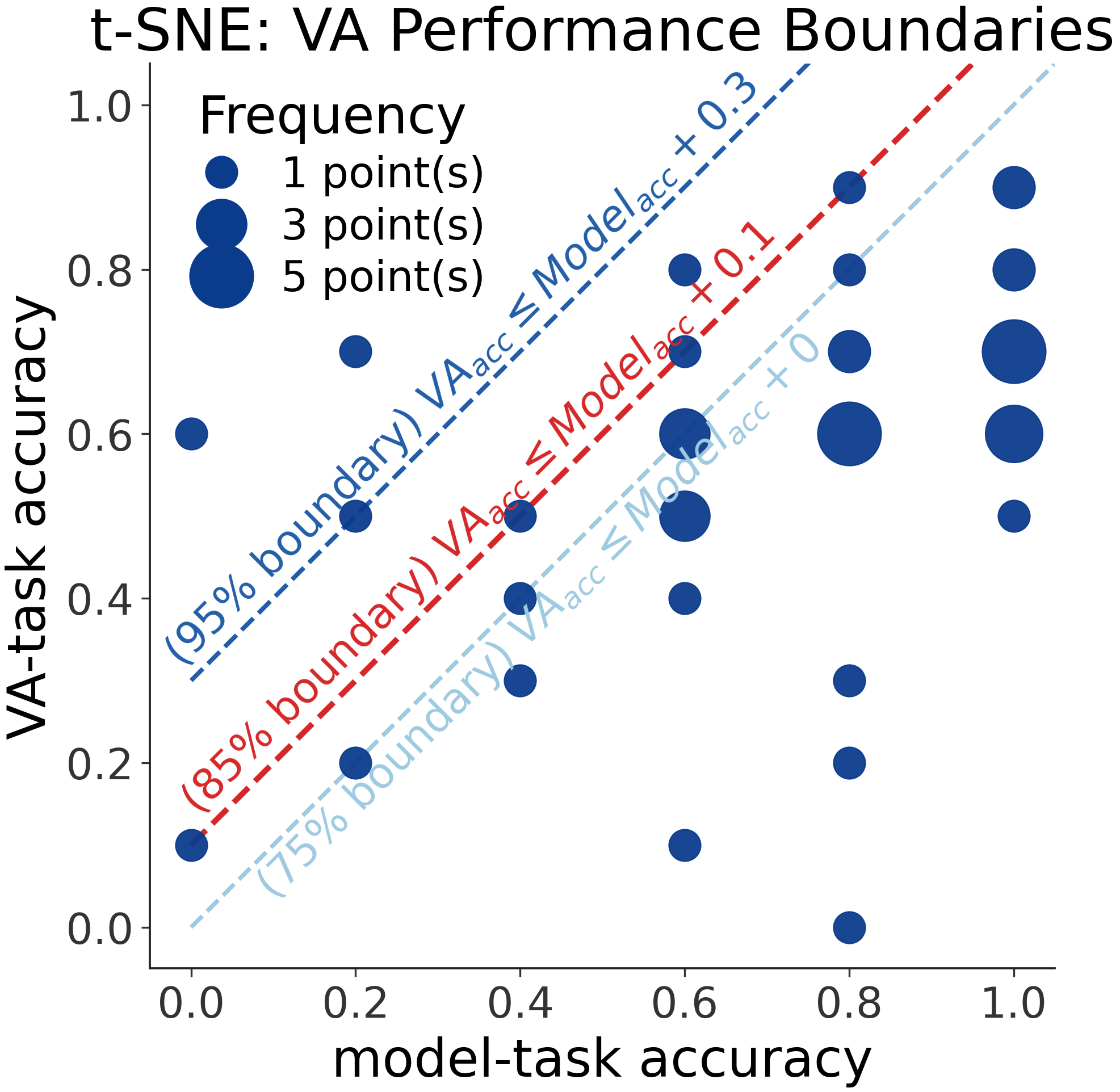}
\caption{Relations between model-task accuracy and VA-task accuracy under PCA and t-SNE. Dashed lines show the strict boundary $VA_{acc}\leq Model_{acc}$ and relaxed boundaries with different margins. Marker size indicates the number of participants sharing the same accuracy pair. Counting marker frequencies, the strict boundary covers 36 of 44 participants (81.8\%) and the $+0.1$ boundary covers 40 of 44 participants (90.9\%) under each model.}
\vspace{-16pt}
\label{fig:md_boundary}
\end{figure}

The clearest pattern supporting H1 appears in VA effectiveness.
As shown in Figure~\ref{fig:md_boundary}, the Pearson correlation between model-task accuracy and VA-task accuracy is $0.653$ for PCA ($p<.001$, 95\% CI $[0.442,0.796]$) and $0.450$ for t-SNE ($p=.002$, 95\% CI $[0.177,0.659]$).
The corresponding Spearman correlations are also positive and significant for PCA ($\rho=0.668$, $p<.001$) and t-SNE ($\rho=0.512$, $p<.001$), indicating that the result does not depend only on the linear-correlation assumption.
These results jointly indicate that participants with higher model-task accuracy also tend to achieve higher downstream VA-task accuracy, even though the strength of the association differs across models.

As a robustness check, we residualize model-task accuracy and VA-task accuracy by whether participants received the pre-questionnaire preparation notice.
The partial correlations remain $0.658$ for PCA ($p<.001$) and $0.436$ for t-SNE ($p=.003$).
Thus, the positive model-task--VA-task accuracy relation is not explained solely by whether participants received the preparation notice.

The boundary pattern provides a complementary descriptive view.
Counting the participant frequencies represented by the markers, the strict relation $VA_{acc}\leq Model_{acc}$ covers 36 of the 44 participants ($81.8\%$) under each model.
When the boundary is relaxed to $VA_{acc}\leq Model_{acc}+0.1$, the coverage rises to 40 of 44 participants ($90.9\%$) for both PCA and t-SNE.
This pattern further illustrates that high VA-task accuracy is rarely observed without comparable model-task accuracy in the current sample.

The efficiency evidence for H1 is mixed when efficiency is operationalized as shorter VA-task completion time.
The relation between model-task accuracy and mean VA-task completion time is moderately positive under PCA ($r=0.387$, $p=.010$; $\rho=0.358$, $p=.017$) and close to zero under t-SNE ($r=0.111$, $p=.475$; $\rho=0.160$, $p=.299$).
The direct time--accuracy relation shows a compatible pattern: under PCA, longer mean VA-task completion time is associated with higher VA-task accuracy ($r=0.494$, $p<.001$; $\rho=0.423$, $p=.004$), whereas the corresponding relation is weaker under t-SNE ($r=0.279$, $p=.067$).
Thus, model literacy appears to support the quality of VA judgments more consistently than the speed of those judgments.
The accuracy results support the effectiveness component of H1, whereas the completion-time patterns suggest that model knowledge may sometimes be invested in more effortful reasoning rather than uniformly faster analysis.

\bsubsec{Evidence for H2: Model Intuitiveness and Knowledge Translation}{H2Results}

Following the model-intuitiveness contrast described in \rsubsec{Def}, PCA is treated as relatively less intuitive and t-SNE as relatively more intuitive in the multidimensional-data experiment.

Building on H1, H2 examines whether model intuitiveness moderates the translation from model literacy to VA performance.
The primary H2 evidence is the contrast between the model-task--VA-task accuracy correlations: the relation is stronger in the less intuitive PCA condition than in the more intuitive t-SNE condition ($r=0.653$ vs.~$r=0.450$).
Steiger's test indicates a significant difference between the two dependent correlations ($\Delta r=0.203$, $p=.027$), indicating that, in the current task design, VA-task accuracy is more strongly associated with model-task accuracy under PCA than under t-SNE.
This contrast is consistent with the proposed role of relative model intuitiveness.
This pattern suggests that, when the relation between model output and visual judgment is less directly readable from the visualization, accurate VA judgment depends more strongly on explicit model knowledge.
Conversely, when the visualized model structure aligns more directly with common spatial reasoning, the dependence on measured model literacy can become looser.
This contrast does not coincide with a large average-accuracy difference between the two model blocks: mean model-task accuracy is similar for PCA and t-SNE ($0.677$ vs.~$0.714$, $p=.538$), as is mean VA-task accuracy ($0.577$ vs.~$0.568$, $p=.518$).
The completion-time comparison provides process context rather than a separate H2 test: the less intuitive PCA condition takes longer on average ($67.1$~s vs.~$35.8$~s, $\Delta=31.3$~s, $p<.001$), which may be consistent with a more effortful translation from model knowledge to VA judgments in settings where model output is less directly readable.

\begin{figure}[t]
\centering
\includegraphics[width=0.48\columnwidth]{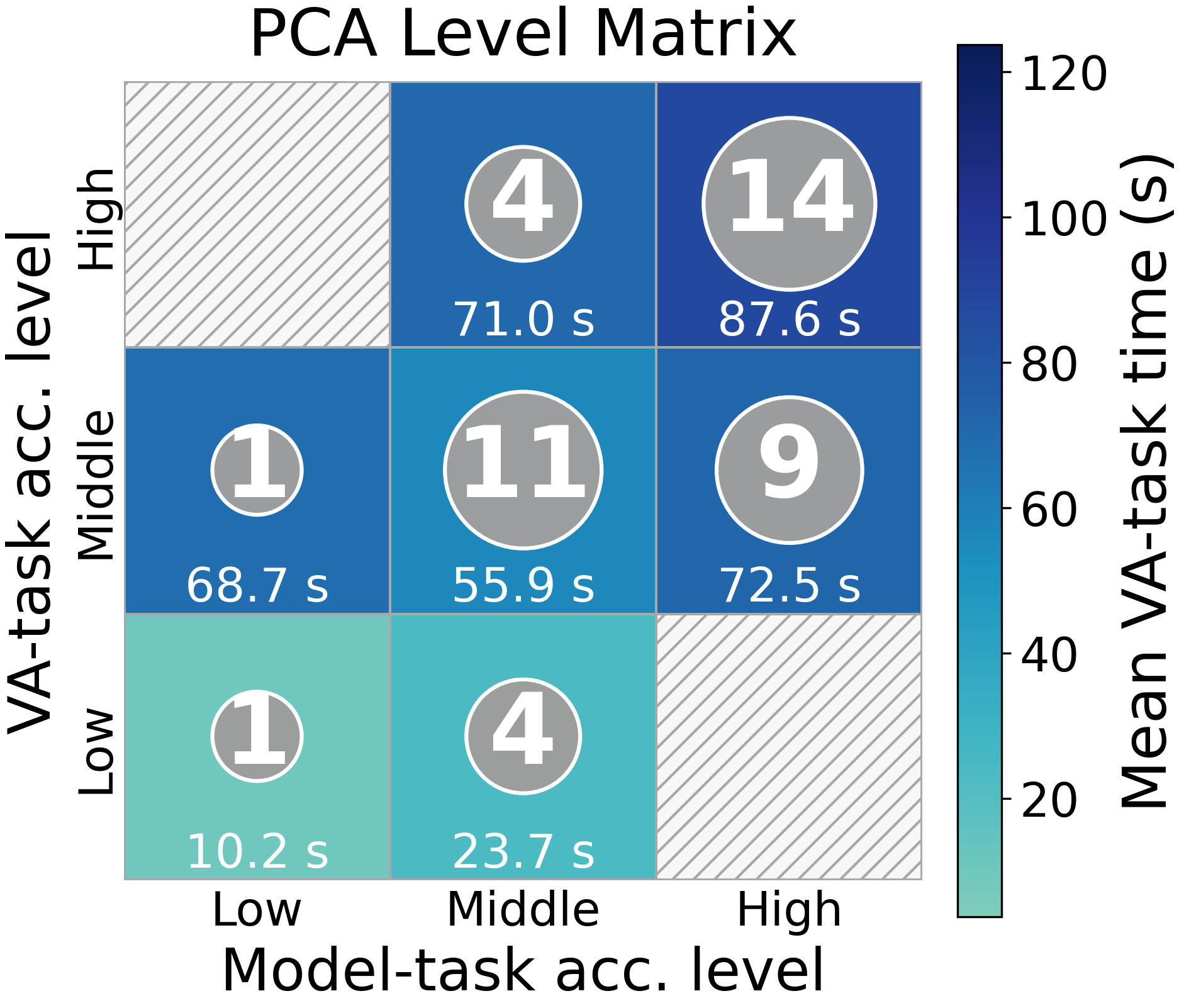}\hfill
\includegraphics[width=0.48\columnwidth]{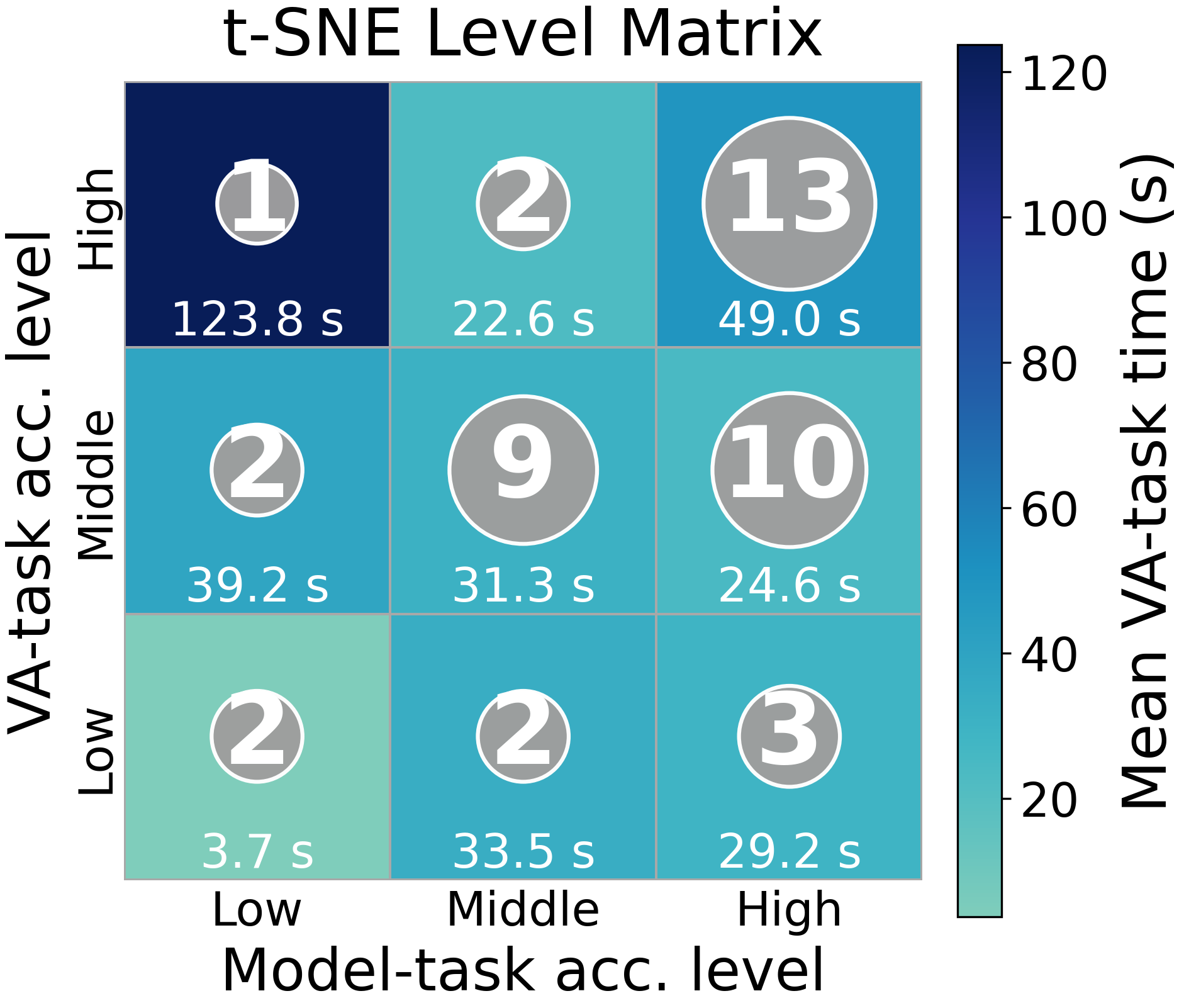}
\caption{Aggregated relations between model-task accuracy and VA-task accuracy under PCA and t-SNE. Low, Middle, and High levels correspond to accuracies $<0.33$, between $0.33$ and $0.66$, and $>0.66$. Circle size indicates participant count, and background color indicates mean VA-task completion time.}
\vspace{-16pt}
\label{fig:md_level_matrix}
\end{figure}

Figure~\ref{fig:md_level_matrix} makes this model-intuitiveness pattern more concrete.
In the less intuitive condition, VA-task accuracy changes more consistently with model-task accuracy: under PCA, the two extreme cross-level cases are absent, with neither a Low model-task accuracy / High VA-task accuracy case nor a High model-task accuracy / Low VA-task accuracy case.
In the more intuitive condition, the mapping is less orderly: under t-SNE, both types still appear, with one case in the former category and three in the latter.
The time pattern in the grouped matrix reinforces this difference.
Under PCA, within the Middle model-task accuracy condition, Low, Middle, and High VA-task accuracy correspond to 23.7~s, 55.9~s, and 71.0~s, respectively; under the High model-task accuracy condition, Middle and High VA-task accuracy correspond to 72.5~s and 87.6~s.
Under t-SNE, by contrast, the corresponding ordering is less regular: within the Middle model-task accuracy condition, High VA-task accuracy is associated with a shorter mean VA-task completion time than both Middle and Low.
Together, the matrix suggests that the pattern under the less directly readable condition is consistent with a more staged translation from model knowledge to VA judgment, whereas a more intuitive representation leaves more room for visible structure to support judgment without the same level of explicit model knowledge.

Overall, the observed pattern is consistent with H2 in the current task design.
More generally, model intuitiveness may help explain how strongly model literacy constrains VA performance.
When the relation between model output and visual judgment is less direct, users need stronger model knowledge and more effortful reasoning to reach accurate VA judgments.
When the visual representation of model output is exposed in a form aligned with users' spatial expectations, VA judgments can depend less tightly on explicit model knowledge.
In this study, PCA and t-SNE instantiate these two patterns, respectively.


\bsec{Discussion}{Discussion}
We discuss implications for designing VA tools and limitations of our study.

\bsubsec{Implications for Designing VA Systems}{D1}




We highlight the significance of the model literacy factor for visual analytics in the summative evaluation stage after implementing a VA system.
In fact, considering the model literacy factor can also improve the design of VA before its implementation.
Designers should be aware of the target user's literacy regarding the specific model embedded in the VA pipeline.
There is a fundamental trade-off between model sophistication and the breadth of the user base a VA system can best serve.
For example, a VA technique designed for a broad population may need to involve simple or familiar data analysis models, or provide sufficient guidance for models unfamiliar to the target users.
In contrast, a VA system customized for a small group of domain specialists can be empowered by sophisticated data analysis models if those models are part of the group's everyday workflow.

Recognizing the significance of model literacy also suggests ways to democratize VA techniques.
A central design target is to make the relationships between model output and VA judgments more readable.
Designers can pursue this target either by choosing data analysis models whose outputs align with users' spatial or conceptual expectations, or by adding visual scaffolds, such as annotations, examples, legends, or interface-level guidance, that explain how model outputs should be interpreted.
Model intuitiveness is not an intrinsic property of an algorithm alone; it also depends on users’ prior knowledge, the analytical task, and how the model output is presented.
Further understanding the relationship between model literacy and downstream VA performance, mediated by the choice of visualization, may help make model-driven VA systems more accessible to broader users.

For VA evaluation, this also means that researchers should report not only participants' general domain or visualization background, but also whether their understanding of the embedded data analysis model is assumed, measured, or supported by the interface.
Otherwise, performance differences in model-driven VA may be incorrectly attributed only to visual encodings, interactions, or interface usability.

\bsubsec{Limitations}{D3}


In this work, we study model literacy in the context of multidimensional data analysis with two common dimensionality-reduction models, PCA and t-SNE.
This focused setting reduces variation attributable to data familiarity and basic scatterplot reading, but it also limits how far the results can be generalized.
Future work should extend the study to more scenarios with complex data types and advanced visualizations, and should examine the interaction among users' data familiarity, visualization literacy, and model literacy with respect to VA performance.

Meanwhile, model literacy in our study is mainly inferred from users' quantitative responses to knowledge-based model-task questions.
It could also be beneficial to collect qualitative evidence to assess model literacy, such as the notes users take when completing VA tasks that require model-related reasoning.
Evaluating model literacy from multiple perspectives may help clarify the underlying mechanism.
Some interpretations, especially those based on the difference between PCA and t-SNE, exceptional cases, or completion-time patterns, should therefore be treated with caution.

The retained responses were collected with and without a pre-questionnaire preparation notice.
A subset of participants was asked to study or review PCA and t-SNE independently before the questionnaire and could choose whether and how to review using self-selected resources, whereas the remaining participants received no such preparation notice and answered based on their existing knowledge.
Actual preparation behavior, duration, and learning resources were not standardized or logged.
The preparation-notice-adjusted analysis preserves the direction and magnitude of the main model-task--VA-task accuracy relations, but it controls only whether the notice was given rather than the amount or quality of preparation.
A confirmatory study should therefore use a homogeneous preparation protocol or explicitly measure and preregister participants' preparation behavior.

Finally, the model tasks always precede the corresponding VA tasks.
This ordering allows model-task accuracy to be observed before downstream VA performance, but answering the model tasks may also activate or reinforce some of the relevant concepts.
Future studies could separate the model assessment and VA-task sessions, include a no-model-task control group, or counterbalance the measurement order.



\bsec{Conclusion}{Conc}

In this work, we introduce a new evaluation factor, namely model literacy, to the summative evaluation of VA systems.
The factor captures users' knowledge of the underlying data analysis model in a VA system.
In the current study, we operationalize model literacy through model-task accuracy and examine its relationship with VA performance in a multidimensional-data setting with PCA and t-SNE as two model-intuitiveness conditions.
The results lead to two main findings.
First, model-task accuracy is positively associated with VA-task accuracy under both models, showing that model literacy supports VA effectiveness.
The completion-time pattern is more complex: higher model-task accuracy does not consistently translate into shorter task completion time.
Instead, completion time appears to reflect not only efficiency but also the reasoning effort required to apply model knowledge, especially when the model requires users to connect abstract model components with visual evidence.
Second, the strength of this relationship varies with model intuitiveness: VA performance is more tightly constrained by model literacy when the relation between model output and visual judgment is less directly readable, and becomes relatively less constrained when the visual representation of model output better matches users' spatial expectations, as observed in the PCA/t-SNE contrast.

Our findings suggest that model literacy may be useful to consider in the summative evaluation of model-driven VA systems and provide directions for studying this factor in broader data, model, and visualization settings.

Future work should extend the current evaluation design beyond dimensionality-reduction models.
One planned direction is to study more types of data.
Such work can further examine whether the observed relations extend to other positions in the VA pipeline while using an independent visualization-literacy measure, broader model-literacy item pools, and preregistered analysis plans.


\bibliographystyle{abbrv-doi}

\bibliography{VALit_VIS}
\clearpage
\end{document}